\documentclass[a4paper]{article}

\usepackage[english]{babel}
\usepackage[utf8x]{inputenc}
\usepackage[T1]{fontenc}
\usepackage[a4paper,top=3cm,bottom=2cm,left=3cm,right=3cm,marginparwidth=1.75cm]{geometry}
\usepackage[english]{babel}
\usepackage{amsmath}
\usepackage{aas_macros}
\usepackage{graphicx}
\usepackage{natbib}
\usepackage[colorinlistoftodos]{todonotes}
\usepackage[colorlinks=true, allcolors=blue]{hyperref}
\usepackage{authblk}
\title{Microlensing planets in the light of the second data release of Gaia}
\author[1]{K.-S. Nikolaus \& M. Hundertmark}
\affil[1]{Astronomisches Rechen-Institut, Zentrum f{\"u}r Astronomie der Universit{\"a}t Heidelberg (ZAH), 69120 Heidelberg, Germany}

\begin{document}
\maketitle

\begin{abstract}
Extrasolar planets found by gravitational microlensing often require
assumptions on the source star distance and relative proper motion. 
Only in a few cases has it been possible to confirm these findings with
space-based observations or high-resolution follow-up. 20 planetary microlensing events can be positionally cross-matched with the second Gaia data release containing parallax and proper motion measurements.

In this work we subject all microlensing planets listed in NASA's Exoplanet Archive 
to an initial consistency check by comparing them with Gaia data release 2 measurements. The resulting list is supposed to serve as a reference for the observers in the microlensing community.

Gravitational microlensing can constrain the physical parameters lens mass and lens distance based on fit parameters, such as the event timescale, the microlensing parallax, and the source star crossing time. If some of these parameters are not available, one needs to resort to indirect means of assessing the events, often involving a Galactic model. In this work, we seek to make an initial assessment of those parameters solely based on Gaia DR2.

We find that 19 of 20 planetary events are consistent within $2\,\sigma$ of their published lens and source distances, whereas 9 of them agree with their published source magnitude within 0.2\,mag. The only event that does not seem to be compatible with the reported distance, is well-constrained and provides plausible distance estimates.
\end{abstract}

\section{Introduction}

Gravitational microlensing is arguably the most successful detection technique for finding extrasolar planets beyond the snowline. Moreover, microlensing planets are located several kiloparsecs away \citep{par06, ben96, wam97} and provide an independent measure of the planet abundance in our Galaxy \citep{sno04,gau02,gou10,sum10,sum11,cas12,suz16,tsa16}. Microlensing planets are detected because light from a distant source star is attracted by a foreground lens star. As a consequence, more light from the source reaches an observer on Earth and the source appears to be brighter. The timescale of the event is related to the total mass of the lens, but also depends on relative proper motion as well as distance to the lens and distance to the source. Some of these parameters can be constrained by analyzing light curves of microlensing events. Further insight can be gained if the distance to the source and its proper motion are known, which is typically achieved by combining a fit to multi-band photometry. Only $\sim1$ in a million stars in the Galactic bulge is sufficiently aligned with a lens star to be detected by the observer as predicted by the seminal paper of \cite{pac86} and the planet name reflects that most of the underlying microlensing events have been discovered by the OGLE \citep{uda94,uda15} and MOA teams \citep{bon04}. 

The ESA Gaia mission \citep{gai16} is obtaining accurate parallaxes and proper motions of about 1.7 billion sources brighter than $G\approx21$. The second data release (Gaia DR2), including five-parameter astrometric solutions with parallaxes, and proper motions, was released to the community on 25 April 2018  \citep{gai18a}. Comparing the list of 58 microlensing planets on NASA's Exoplanet Archive\footnote{\url{http://exoplanetarchive.ipac.caltech.edu} 25 April 2018} with the initial Gaia data release \citep{gai16b} reveals that 13 microlensing events can be positionally cross-matched and all of them can be found in Gaia DR2, albeit not with five-parameter astrometric solutions. In the following we describe which observable microlensing parameters are related to Gaia parameters and suggest a way to check if they comply with findings of Gaia. 

Most microlensing events follow a simple symmetric Paczy{\'n}ski light curve. In a co-linear lens-source-observer configuration the source star image appears as Einstein ring radius
\begin{equation}
\theta_{\mathrm{E}} = \sqrt{\frac{4 G M_{\mathrm{L}}}{c^2} \left(D^{-1}_{\mathrm{L}}-D^{-1}_{\mathrm{S}}\right)}
\label{eq1}
\end{equation} 
constraining the typical angular scale of the effect, where $D_{\mathrm{L}}$ denotes the distance from observer to deflecting lens of mass $M_{\mathrm{L}}$ and $D_{\mathrm{S}}$ denote the distance to the source star.

The size of the Einstein radius is on the order of $\sim1\,\mathrm{mas}$ and thus cannot be resolved. The relative proper motion between lens and source
\begin{equation}
\mu_{\mathrm{rel}} = \mu_{\mathrm{S}} - \mu_{\mathrm{L}},
\label{eq2}
\end{equation} 
changes the alignment of the lens and source stars. This causes the brightness to change accordingly since Einstein's deflection angle depends on the impact parameter and thus magnifies the source. The typical time-scale of the event can be expressed as
\begin{equation}
t_{\mathrm{E}} = \frac{\theta_{\mathrm{E}}}{\mu_{\mathrm{rel}}}.
\label{eq3}
\end{equation}
The former is called Einstein time and can be obtained from a fit to the microlensing light curve. Usually, the source star is sufficiently bright to be seen when being magnified. If the source star is sufficiently bright, one can expect to obtain a parallax vector constraining $D_{\mathrm{S}}$ and the proper motion $\mu_{\mathrm{S}}$ which would leave us with the task to find a constraint on $\mu_{\mathrm{L}}$. 

It should be stated that some microlensing light curves themselves provide a way of measuring $\mu_{\mathrm{rel}}$ by using finite source effects, namely the angular source star radius $\rho$ expressed in units of $\theta_{\mathrm{E}}$ so that 
\begin{equation}
\theta_{\mathrm{E}} = \frac{\theta_{\mathrm{\star}}}{\rho},
\label{eq4}
\end{equation}
where $\theta_{\mathrm{\star}}$ is the angular size of the source star inferred from the source color and $\rho$ can be retrieved from a fit. Using Eq.~\ref{eq3} then leads to the relative proper motion. A direct comparison with Gaia DR2 is hard to achieve because the lens is usually too faint to be detected. Within the scope of this work, we could only check if the proper motion of the source or the lens star exceeds the expected distribution of $\mu_{\mathrm{L}}$ or $\mu_{\mathrm{S}}$ that can be obtained from Gaia DR2 itself. 

Finally, an asymmetry in the light curve can lead to the detection of a parallax vector which we will refer to as microlensing parallax in order to distinguish it from the parallax measured by Gaia. Observing a microlensing event from different observatories, at different times of the year and/or by using satellite observations \citep{gou94, gou00} introduces the microlensing parallax as further fit parameter. It is related to source and lens distance as well as the Einstein radius $\theta_{\mathrm{E}}$ through
\begin{equation}
\pi_{\mathrm{E}} = \frac{1}{\theta_{\mathrm{E}}} \left(\frac{\mathrm{AU}}{D_{\mathrm{L}}} - \frac{\mathrm{AU}}{D_{\mathrm{S}}}\right).
\label{eq5}
\end{equation} 

\section{Rationale of the comparison}

At first glance, the comparison of existing microlensing planets reported in the literature with Gaia DR2 data seems to be straight-forward. One needs to cross-check if the source star is in the Gaia catalog, ensure that the reported brightness is consistent and apply the corrected parameters to the reported physical parameter estimates. In practice, one faces several challenges as far as selecting a meaningful sample is concerned. Not all published planets come with reported uncertainties on all relevant parameters. In some cases, the parameter space is too complicated to provide an unambiguously set of fit parameters. Moreover, when uncertainties are reported, it is often not clear how the underlying parameter estimates are distributed and if the parameters for the best solution are consistent. In this work, we will follow a simple approach and rely on the best fit in a least-squares sense and use the corresponding parameters as a starting point. 

\subsection{Initial selection of planets}

First we devise a filter criterion to determine which planets can be used for resampling the lens mass distribution. Out of 58 planets listed as confirmed in NASA's Exoplanet Archive, we only consider 53 due to a lack of reported values for $\theta_{\mathrm{E}}$. Therefore, the microlensing events MOA-2007-BLG-192L \citep{2008ApJ...684..663B} and OGLE-2016-BLG-0263L \citep{2017AJ....154..133H} are missing.

If applicable, we also compare $\mu_{\mathrm{rel}}$, $\pi_{\mathrm{E}}$, $t_{\mathrm{E}}$, $\theta_{\star}$ and $\rho_{\star}$ using Eqs.~\ref{eq1}, \ref{eq3}, \ref{eq4} and \ref{eq5} in order to assess whether they describe our event and to perform a consistency check so one can argue how compatible they are with the reported $\theta_{\mathrm{E}}$, $D_{\mathrm{L}}$ and $D_{\mathrm{S}}$. Unmentioned source distances are assumed to be $D_{\mathrm{S}}=8\,\mathrm{kpc}$. Moreover, the given mass ratio
\begin{equation}
q=\frac{M_{\mathrm{pl}}}{M_{\mathrm{host}}}
\end{equation}
is compared with the reported planet and host star masses. The planet is then obtained from
\begin{equation}
M_{\mathrm{pl}}=\frac{q}{q+1}\cdot\frac{\theta^2_{\mathrm{E}}}{\kappa}\cdot \left(\frac{\mathrm{AU}}{D_{\mathrm{L}}} - \frac{\mathrm{AU}}{D_{\mathrm{S}}}\right)^{-1}.
\label{eq6}
\end{equation} 
Depending on the available parameters we determine the following quantities where applicable:

\begin{equation}
M_{\mathrm{pl}}=\frac{q}{q+1}\cdot\frac{\theta^2_{\mathrm{E}}}{\kappa\cdot\pi_{\mathrm{E}}},
\label{eq7}
\end{equation} 
\begin{equation}
M_{\mathrm{pl}}=\frac{q}{q+1}\cdot\frac{(\mu_{\mathrm{rel}}\cdot t_{\mathrm{E}})^2}{\kappa}\cdot \left(\frac{\mathrm{AU}}{D_{\mathrm{L}}} - \frac{\mathrm{AU}}{D_{\mathrm{S}}}\right)^{-1},
\label{eq8}
\end{equation} 

\begin{equation}
M_{\mathrm{pl}}=\frac{q}{q+1}\cdot\frac{(\mu_{\mathrm{rel}}\cdot t_{\mathrm{E}})^2}{\kappa\cdot\pi_{\mathrm{E}}},
\label{eq9}
\end{equation} 
\begin{equation}
M_{\mathrm{pl}}=\frac{q}{q+1}\cdot\frac{\theta_{\star}}{\kappa\cdot\rho}\cdot \left(\frac{\mathrm{AU}}{D_{\mathrm{L}}} - \frac{\mathrm{AU}}{D_{\mathrm{S}}}\right)^{-1},
\label{eq10}
\end{equation} 
\begin{equation}
M_{\mathrm{pl}}=\frac{q}{q+1}\cdot\frac{\theta_{\star}}{\kappa\cdot\rho\cdot\pi_{\mathrm{E}}}.
\label{eq11}
\end{equation}

We accept a discrepancy of 20\,\% on $\theta_{\mathrm{E}}$ and 10\,\% on the other aforementioned parameter values as a lower threshold. Any other preselection would drastically decrease the number of considerable planets. If these parameters are completely incompatible, we refrain from including them in our study. This applies to the events MOA-2010-BLG-328L \citep{2013ApJ...779...91F}, OGLE-2005-BLG-169L \citep{2006ApJ...644L..37G}, and OGLE-2013-BLG-0341L B \citep{2014Sci...345...46G}.

After passing these checks, all available parameters are used to calculate $M_{\mathrm{pl}}$ based on Eqs.~\ref{eq6} to~\ref{eq11}, where $\kappa=8.144 \frac{\mathrm{mas}}{\mathrm{M}_{\odot}}$. The results are compared with the reported planet masses and the method delivering the closest deviation is chosen to be used. It turns out, that in many cases applying Eq.~\ref{eq6} leads to the most reliable results as compared to the published planet mass whereas the other approaches are just partially applicable.

\begin{figure}
\centering
  \resizebox{.8\textwidth}{!}{\includegraphics{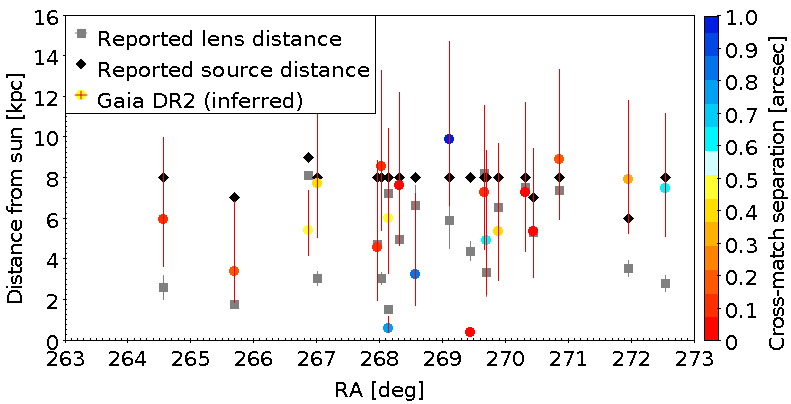}}
  \caption{Distances to the reported microlensing planets and their respective host stars is shown along with the corresponding cross-matched distances based on Gaia DR2 parallaxes. The sample is limited to microlensing events towards the Galactic center.}
  \label{fig1}
\end{figure}

For the actual comparison we need a five-parameter astrometric solution including parallaxes. That reduces our target list to 20 microlensing planets, and some are listed with negative parallaxes, which is not a surprise given that most microlensing events are located in crowded fields and the limiting magnitude is $G\approx18$. In addition to the positional cross-match we are checking if the reported I magnitudes are consistent with the blue and red Gaia magnitudes $G_{\mathrm{BP}}, G_{\mathrm{RP}}$ which have been transformed to the Johnson-Cousin I band using the relations of \cite{jor10}. Fig.~\ref{fig1} shows the distance to lens and source as well as the cross-matched Gaia DR2 candidate

\section{Results}

We find that 9 cross-matched stars are within 0.5\,arcsec of the reported target position and within 0.2\,mag of the reported source magnitude. Two more events are positionally cross-matched within 0.5\,arcsec, but do not match the source magnitude. Both events are highly blended, but the inferred distance cannot confirm or rule out if the lens is the blend. 

There was only one cross-matched planetary event with consistent brightness that did not match the reported lens and source distance. \cite{2015ApJ...804...33S} report for OGLE-2011-BLG-0265L a source color $V-I$ of 3.2 which differs from the Gaia DR2 target. The Gaia target is likely not the blend, because it is reported it to be $>20$\,mag in the I band. The cross-match separation is within 0.06\,arcsec. The magnitude of the relative proper motion of the Gaia DR2 target is 10\,mas/yr. Typical values for the relative proper motion are in the range of 2 and 8\,mas/yr which by itself would not exclude the possibility of the Gaia DR2 target being the source. Since the event was discovered, the position should not have changed enough to affect the cross-match. We can also exclude that the deviation comes from the inferred distances, since a direct inversion of the parallax is similarly close. Due to the faintness of the event in Gaia ($G\approx 18.9$) we expect that the crowded field has affected the event.  

As a side-remark, Gaia DR2 also reports duplicated sources as diagnostic information which can indicate highly blended events before they occur. In the given selection OGLE-2006-BLG-109L \citep{2010ApJ...713..837} and MOA-2008-BLG-379L \citep{2014ApJ...780..123S} were reported and duplicated sources.

Considering different avenues of comparing parameters, we compare in Table~\ref{tab1} the reported lens and source distance with the distance of the nearest catalog entry based on the discovery paper\footnote{For some of the events revised or extended parameter estimates are available \citep{bat17, bat14, ben06, don09, ben15, ben10, bea16, tsa14}}. The reference distance and asymmetric uncertainties are based on the inferred distances provided by \cite{bai18} because some of the cross-matched targets are reported to have negative parallaxes. That increases our sample to 20 microlensing planets. The approach has already been tested on Gaia DR1 data \citep{bai15, ast16a, ast16b}. We report if $D_{\mathrm{S}}$ or $D_{\mathrm{L}}$ are within $2\,\sigma$ of the respective asymmetric uncertainties. 

\begin{table}[h]
\caption{Comparison between the reported $D_{\mathrm{S}}, D_{\mathrm{L}}$ in the discovery paper and the inferred distance by \cite{bai18}. Converted Johnson-Cousins magnitudes and colors $I_{\mathrm{G}},(V-I)_{\mathrm{G}}$ are calculated based on \citep{jor10}. The first part of the table contains matches within 0.2\,mag of the published source star and a separation below 0.5\,arcec. The second part contains events that cannot be matched to the source magnitude. The last part contains entries without Gaia colors and separations $>0.5$\,arcsec.}              
\label{table:1}      
\centering 
\renewcommand{\arraystretch}{1.3}
\begin{tabular}{c c c c c}          
\hline\hline                        
Hostname & $\Delta D_{\mathrm{S}}$ & $\Delta D_{\mathrm{L}}$ & $I_{\mathrm{G}}$ & $(V-I)_{\mathrm{G}}$ \\ 
 & $\in [\pm 2\sigma]$ & $\in [\pm 2\sigma]$ & $[\mathrm{mag}]$  & $[\mathrm{mag}]$\\
\hline     
MOA-2009-BLG-266L & yes & yes & 15.9 & 1.7 \\
MOA-2011-BLG-028L & yes & yes & 15.3 & 1.8 \\
OGLE-2008-BLG-092L & yes & yes & 13.9 & 2.0 \\
OGLE-2011-BLG-0265L & no & no & 17.6 & 2.4 \\
OGLE-2012-BLG-0358L & yes & yes & 16.5 & 2.4 \\
OGLE-2013-BLG-0102L & yes & yes & 17.3 & 2.5 \\
OGLE-2015-BLG-0051L & yes & yes & 16.8 & 2.3 \\
OGLE-2016-BLG-0263L & yes & yes & 17.0 & 2.0 \\
OGLE-2017-BLG-1522L & yes & yes & 17.2 & 1.7 \\
\hline
OGLE-2011-BLG-0251L & yes & yes & 15.8 & 3.0 \\
MOA-2010-BLG-117L & yes & yes & 16.8 & 2.0 \\
\hline
MOA-2010-BLG-073L & yes & yes & 15.4 & 2.1 \\
MOA-2012-BLG-505L & yes & yes & 17.0 & 2.2 \\
OGLE-2006-BLG-109L & yes & no & 16.8 & 2.2 \\
OGLE-2007-BLG-368L & yes & yes & 16.0 & 2.0 \\
MOA-2008-BLG-379L & yes & yes & -- & -- \\
MOA-2012-BLG-006L & yes & yes & -- & -- \\
OGLE-2005-BLG-390L & yes & yes & -- & -- \\
OGLE-2012-BLG-0406L & yes & yes & -- & -- \\
OGLE-2017-BLG-0173L & yes & yes & -- & -- \\
\hline                                          
\end{tabular}
\label{tab1}
\end{table}

\section{Conclusions}

\begin{figure}
\centering
  \resizebox{.5\textwidth}{!}{\includegraphics{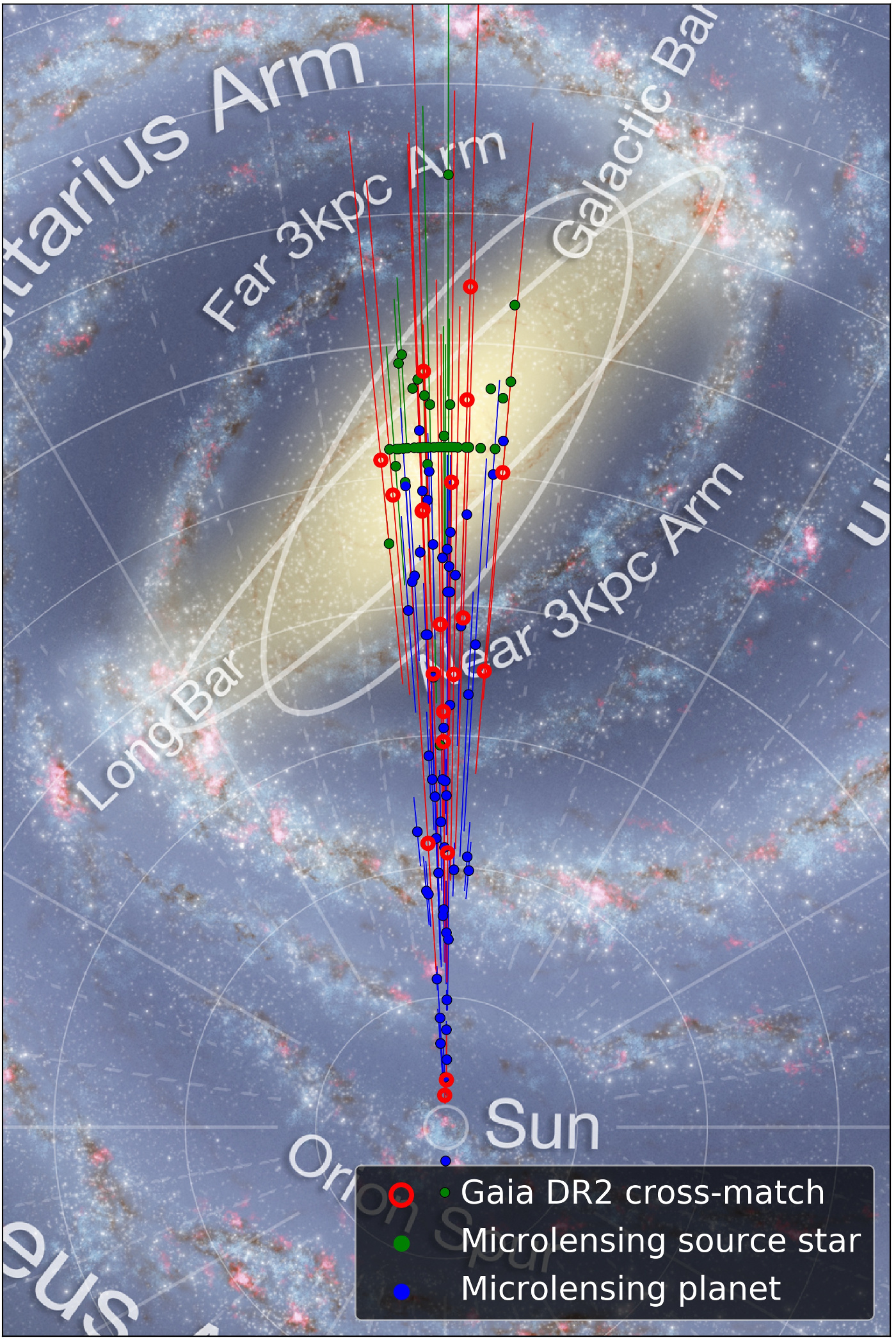}}
  \caption{Positions of source stars and reported microlensing planets are shown in an artist's impression of our Galaxy along with cross-matched Gaia DR2 distances. Credit:
NASA/JPL-Caltech/ESO/R. Hurt}
  \label{fig2}
\end{figure}

We have subjected the published microlensing planets to a first compatibility check with Gaia DR2. We have shown that 19 of 20 planetary events with a five-parameter astrometric solution are within $2\,\sigma$ of the reported source distance or the expected lens distance and 9 of them can be matched to the source magnitude within 0.2\,mag.

When revisiting the published parameters, we find that the lack of uncertainties, parameters and Monte C samples makes a fair comparison difficult. We would like to note that we have also treated all events equally, but feature-rich and long events constraining source radius and parallax provide estimates that are independent of a Galactic model and by definition are more informative.

The lens and source star distances of microlensing events are likely to be constrained in a better way by using color-magnitude diagrams along with source and blend flux as fit parameter. For OGLE-2011-BLG-0265L, where the inferred Gaia distance is not compatible with source or lens, we assume that the corresponding Gaia DR2 catalog entry belongs to the source star.

Since we can not reliably assign a Gaia source identifier to the lens, the source or the blend, the reported proper motion can not be used for further constraining the lens parameters. For our selection of 20 events, the proper motion stays around $7\pm3\,\,\mathrm{mas}/\mathrm{yr}$ and does not require a more careful treatment of the observational epoch. The cross-matched targets for OGLE-2005-BLG-390L and OGLE-2005-BLG-265L have the highest reported proper motion (10.0 and 14.6\,mas/yr respectively).

Fig.~\ref{fig2} illustrates how the assumed microlensing planet distribution looks like. Green circles correspond to the source star positions which are quite often constrained to 8\,kpc. We also want to highlight that the published lens distances are hinting at more lens stars closer to the sun than one would expect. These events are more likely to enable the measurement of $\pi_{\mathrm{E}}$ and might be affected by a publication bias which makes it easier to unambiguously characterize them as a planet.

We expect that the distribution of parameters provided by Gaia DR2 will also contribute decisively to refining the most recent Galactic models \citep{pas18} that can be used to assess the nature of microlensing events as well as complementing high-resolution follow-up observations. It might be useful for the inferred Gaia DR2 distance estimates to compare all single and binary microlensing events with corresponding microlensing parallax measurements in order to see how the inferred distances at several kiloparsecs are affected by crowded fields.

\section*{Acknowledgement}
This work has made use of data from the European Space Agency (ESA) mission
{\it Gaia} (\url{https://www.cosmos.esa.int/gaia}), processed by the {\it Gaia}
Data Processing and Analysis Consortium (DPAC, \url{https://www.cosmos.esa.int/web/gaia/dpac/consortium}). Funding for the DPAC
has been provided by national institutions, in particular the institutions
participating in the {\it Gaia} Multilateral Agreement.

This research has made use of the NASA Exoplanet Archive, which is operated by the California Institute of Technology, under contract with the National Aeronautics and Space Administration under the Exoplanet Exploration Program. 

This work is based in part on services provided by the GAVO data center.

\begin{table}[h]
\caption{Selection of 50 planets listed in NASA's Exoplanet Archive.}              
\label{table:2}      
\centering 
\renewcommand{\arraystretch}{1.08}
\begin{tabular}{c c c c c c}          
\hline\hline                        
\hspace{-0.9cm}Eventname & \hspace{-0.4cm}$M_{\mathrm{planet}}$ $[\mathrm{M_{\oplus}]}$ & \hspace{-0.4cm}$D_{\mathrm{L}}$ [kpc] & $\theta_{\mathrm{E}}$ [mas] & $\pi_{\mathrm{E}}$ & Reference\\    
\hline     

\hspace{-0.9cm} MOA-2007-BLG-400 & \hspace{-0.4cm}$ 217.0 ^{+ 152.0 } _{- 79.0 }$ & \hspace{-0.4cm} 5.8 $\pm$ 0.7 & 0.3191 $\pm$ 0.0171 & --- $\pm$ --- & \cite {2009ApJ...698.1826D} \\
\hspace{-0.9cm} MOA-2008-BLG-310 & \hspace{-0.4cm}$ 21.5 ^{+ 52.3 } _{- 12.3 }$ & \hspace{-0.4cm} 7.7 $\pm$ 1.1 & 0.155 $\pm$ 0.011 & --- $\pm$ --- & \cite {2010ApJ...711..731J} \\
\hspace{-0.9cm} MOA-2008-BLG-379 & \hspace{-0.4cm}$ 1160.0 ^{+ 1330.0 } _{- 670.0 }$ & \hspace{-0.4cm} 3.3 $\pm$ 1.25 & 0.88 $\pm$ 0.19 & --- $\pm$ --- & \cite {2014ApJ...780..123S} \\
\hspace{-0.9cm} MOA-2009-BLG-266 & \hspace{-0.4cm}$ 10.44 ^{+ 2.46 } _{- 1.96 }$ & \hspace{-0.4cm} 3.04 $\pm$ 0.33 & 0.98 $\pm$ 0.04 & 0.2 $\pm$ 0.1 & \cite {2011ApJ...741...22M} \\
\hspace{-0.9cm} MOA-2009-BLG-319 & \hspace{-0.4cm}$ 46.0 ^{+ 68.8 } _{- 22.7 }$ & \hspace{-0.4cm} 6.1 $\pm$ 1.15 & 0.34 $\pm$ 0.03 & 0.17 $\pm$ 0.28 & \cite {2011ApJ...728..120M} \\
\hspace{-0.9cm} MOA-2009-BLG-387 & \hspace{-0.4cm}$ 750.0 ^{+ 1630.0 } _{- 470.0 }$ & \hspace{-0.4cm} 5.69 $\pm$ 2.185 & 0.31 $\pm$ 0.03 & 2.69 $\pm$ 0.62 & \cite {2011AA...529A.102B} \\
\hspace{-0.9cm} MOA-2010-BLG-073 & \hspace{-0.4cm}$ 3330.0 ^{+ 1050.0 } _{- 830.0 }$ & \hspace{-0.4cm} 2.8 $\pm$ 0.4 & 0.557 $\pm$ 0.09 & 0.4 $\pm$ 0.1 & \cite {2013ApJ...763...67S} \\
\hspace{-0.9cm} MOA-2010-BLG-353 & \hspace{-0.4cm}$ 44.0 ^{+ 87.6 } _{- 32.3 }$ & \hspace{-0.4cm} 6.43 $\pm$ 1.12 & 0.187 $\pm$ 0.089 & --- $\pm$ --- & \cite {2015MNRAS.454..946R} \\
\hspace{-0.9cm} MOA-2010-BLG-477 & \hspace{-0.4cm}$ 500.0 ^{+ 243.0 } _{- 178.0 }$ & \hspace{-0.4cm} 2.3 $\pm$ 0.6 & 1.38 $\pm$ 0.11 & --- $\pm$ --- & \cite {2012ApJ...754...73B} \\
\hspace{-0.9cm} MOA-2011-BLG-028 & \hspace{-0.4cm}$ 28.7 ^{+ 39.9 } _{- 13.7 }$ & \hspace{-0.4cm} 7.38 $\pm$ 0.57 & 0.337 $\pm$ 0.053 & 0.11 $\pm$ 0.22 & \cite {2016ApJ...820....4S} \\
\hspace{-0.9cm} MOA-2011-BLG-262 & \hspace{-0.4cm}$ 37.5 ^{+ 70.3 } _{- 19.1 }$ & \hspace{-0.4cm} 7.0 $\pm$ 0.95 & 0.205 $\pm$ 0.016 & --- $\pm$ --- & \cite {2014ApJ...785..155B} \\
\hspace{-0.9cm} MOA-2011-BLG-293 & \hspace{-0.4cm}$ 16.1 ^{+ 25.5 } _{- 6.8 }$ & \hspace{-0.4cm} 7.72 $\pm$ 0.44 & 0.26 $\pm$ 0.02 & 2.94 $\pm$ 2.76 & \cite {2012ApJ...755..102Y} \\
\hspace{-0.9cm} MOA-2011-BLG-322 & \hspace{-0.4cm}$ 7100.0 ^{+ 16200.0 } _{- 3800.0 }$ & \hspace{-0.4cm} 7.56 $\pm$ 0.91 & 0.3 $\pm$ 0.01 & --- $\pm$ --- & \cite {2014MNRAS.439..604S} \\
\hspace{-0.9cm} MOA-2012-BLG-006 & \hspace{-0.4cm}$ 2590.0 ^{+ 3010.0 } _{- 1280.0 }$ & \hspace{-0.4cm} 5.3 $\pm$ 1.05 & 0.489 $\pm$ 0.082 & --- $\pm$ --- & \cite {2017AA...604A.103P} \\
\hspace{-0.9cm} MOA-2012-BLG-505 & \hspace{-0.4cm}$ 5.7 ^{+ 13.6 } _{- 3.4 }$ & \hspace{-0.4cm} 7.21 $\pm$ 1.125 & 0.12 $\pm$ 0.02 & --- $\pm$ --- & \cite {2017AJ....154...35N} \\
\hspace{-0.9cm} MOA-2013-BLG-605 & \hspace{-0.4cm}$ 25.2 ^{+ 17.7 } _{- 8.5 }$ & \hspace{-0.4cm} 3.6 $\pm$ 0.7 & 0.48 $\pm$ 0.06 & 0.29 $\pm$ 0.12 & \cite {2016ApJ...825..112S} \\
\hspace{-0.9cm} MOA-2016-BLG-227 & \hspace{-0.4cm}$ 631.0 ^{+ 963.0 } _{- 298.0 }$ & \hspace{-0.4cm} 6.5 $\pm$ 1.0 & 0.227 $\pm$ 0.0075 & 1.3 $\pm$ 2.4 & \cite {2017AJ....154....3K} \\
\hspace{-0.9cm} MOA-bin-1 & \hspace{-0.4cm}$ 1470.0 ^{+ 2780.0 } _{- 910.0 }$ & \hspace{-0.4cm} 5.1 $\pm$ 1.55 & 0.77 $\pm$ 0.11 & --- $\pm$ --- & \cite {2012ApJ...757..119B} \\
\hspace{-0.9cm} OGLE-2003-BLG-235 & \hspace{-0.4cm}$ 717.0 ^{+ 454.0 } _{- 280.0 }$ & \hspace{-0.4cm} 5.8 $\pm$ 0.65 & 0.52 $\pm$ 0.08 & --- $\pm$ --- & \cite {2004ApJ...606L.155B} \\
\hspace{-0.9cm} OGLE-2005-BLG-071 & \hspace{-0.4cm}$ 1030.0 ^{+ 269.0 } _{- 220.0 }$ & \hspace{-0.4cm} 3.2 $\pm$ 0.4 & 0.84 $\pm$ 0.05 & --- $\pm$ --- & \cite {2004ApJ...606L.155B} \\
\hspace{-0.9cm} OGLE-2005-BLG-390 & \hspace{-0.4cm}$ 3.75 ^{+ 4.92 } _{- 1.8 }$ & \hspace{-0.4cm} 6.6 $\pm$ 1.0 & 0.205 $\pm$ 0.0295 & --- $\pm$ --- & \cite {2006Natur.439..437B} \\
\hspace{-0.9cm} OGLE-2006-BLG-109 & \hspace{-0.4cm}$ 234.7 ^{+ 41.6 } _{- 36.8 }$ & \hspace{-0.4cm} 1.51 $\pm$ 0.115 & 1.505 $\pm$ 0.0 & --- $\pm$ --- & \cite {2010ApJ...713..837} \\
\hspace{-0.9cm} OGLE-2006-BLG-109 & \hspace{-0.4cm}$ 87.5 ^{+ 15.6 } _{- 13.7 }$ & \hspace{-0.4cm} 1.51 $\pm$ 0.115 & 1.505 $\pm$ 0.0 & --- $\pm$ --- & \cite {2010ApJ...713..837} \\
\hspace{-0.9cm} OGLE-2007-BLG-349 & \hspace{-0.4cm}$ 79.2 ^{+ 24.1 } _{- 17.5 }$ & \hspace{-0.4cm} 2.76 $\pm$ 0.38 & 1.15 $\pm$ 0.05 & 0.2 $\pm$ 0.1 & \cite {2016AJ....152..125B} \\
\hspace{-0.9cm} OGLE-2007-BLG-368 & \hspace{-0.4cm}$ 21.4 ^{+ 35.2 } _{- 11.5 }$ & \hspace{-0.4cm} 5.9 $\pm$ 1.15 & 0.529 $\pm$ 0.084 & 1.78 $\pm$ 0.28 & \cite {2010ApJ...710.1641S} \\
\hspace{-0.9cm} OGLE-2008-BLG-092 & \hspace{-0.4cm}$ 56.7 ^{+ 14.1 } _{- 12.6 }$ & \hspace{-0.4cm} 8.1 $\pm$ 0.0 & 0.344 $\pm$ 0.02 & --- $\pm$ --- & \cite {2014ApJ...795...42P} \\
\hspace{-0.9cm} OGLE-2008-BLG-355 & \hspace{-0.4cm}$ 1460.0 ^{+ 2920.0 } _{- 800.0 }$ & \hspace{-0.4cm} 6.8 $\pm$ 1.1 & 0.28 $\pm$ 0.03 & --- $\pm$ --- & \cite {2014ApJ...788..128K} \\
\hspace{-0.9cm} OGLE-2011-BLG-0251 & \hspace{-0.4cm}$ 167.8 ^{+ 71.7 } _{- 42.2 }$ & \hspace{-0.4cm} 2.57 $\pm$ 0.61 & 0.749 $\pm$ 0.283 & 0.3 $\pm$ 0.1 & \cite {2013AA...552A..70K} \\
\hspace{-0.9cm} OGLE-2011-BLG-0265 & \hspace{-0.4cm}$ 274.3 ^{+ 98.3 } _{- 73.2 }$ & \hspace{-0.4cm} 4.38 $\pm$ 0.48 & 0.42 $\pm$ 0.04 & 0.24 $\pm$ 0.12 & \cite {2015ApJ...804...33S} \\
\hspace{-0.9cm} OGLE-2012-BLG-0026 & \hspace{-0.4cm}$ 35.5 ^{+ 23.4 } _{- 10.4 }$ & \hspace{-0.4cm} 4.08 $\pm$ 0.38 & 0.91 $\pm$ 0.09 & 0.1 $\pm$ 0.1 & \cite {2013ApJ...762L..28H} \\
\hspace{-0.9cm} OGLE-2012-BLG-0026 & \hspace{-0.4cm}$ 214.0 ^{+ 143.0 } _{- 63.0 }$ & \hspace{-0.4cm} 4.08 $\pm$ 0.38 & 0.91 $\pm$ 0.09 & 0.1 $\pm$ 0.1 & \cite {2013ApJ...762L..28H} \\
\hspace{-0.9cm} OGLE-2012-BLG-0358 & \hspace{-0.4cm}$ 585.0 ^{+ 131.0 } _{- 114.0 }$ & \hspace{-0.4cm} 1.76 $\pm$ 0.13 & 0.29 $\pm$ 0.03 & 1.5 $\pm$ 0.14 & \cite {2013ApJ...762L..28H} \\
\hspace{-0.9cm} OGLE-2012-BLG-0406 & \hspace{-0.4cm}$ 2460.0 ^{+ 1570.0 } _{- 740.0 }$ & \hspace{-0.4cm} 4.97 $\pm$ 0.29 & 0.57 $\pm$ 0.07 & --- $\pm$ --- & \cite {2014ApJ...782...47P} \\
\hspace{-0.9cm} OGLE-2012-BLG-0563 & \hspace{-0.4cm}$ 122.4 ^{+ 82.7 } _{- 65.9 }$ & \hspace{-0.4cm} 1.3 $\pm$ 0.7 & 1.36 $\pm$ 0.13 & 0.62 $\pm$ 0.34 & \cite {2015ApJ...809...74F} \\
\hspace{-0.9cm} OGLE-2012-BLG-0724 & \hspace{-0.4cm}$ 129.0 ^{+ 248.0 } _{- 70.0 }$ & \hspace{-0.4cm} 6.7 $\pm$ 1.15 & 0.239 $\pm$ 0.028 & 8.14 $\pm$ 1.24 & \cite {2016ApJ...824..139H} \\
\hspace{-0.9cm} OGLE-2012-BLG-0950 & \hspace{-0.4cm}$ 44.5 ^{+ 34.2 } _{- 20.1 }$ & \hspace{-0.4cm} 3.0 $\pm$ 0.95 & 1.09 $\pm$ 0.13 & 0.6 $\pm$ 0.5 & \cite {2017AJ....153....1K} \\
\hspace{-0.9cm} OGLE-2013-BLG-0102 & \hspace{-0.4cm}$ 4190.0 ^{+ 1170.0 } _{- 810.0 }$ & \hspace{-0.4cm} 3.04 $\pm$ 0.31 & 0.43 $\pm$ 0.04 & 0.5 $\pm$ 0.1 & \cite {2015ApJ...798..123J} \\
\hspace{-0.9cm} OGLE-2013-BLG-0132 & \hspace{-0.4cm}$ 103.0 ^{+ 122.0 } _{- 56.0 }$ & \hspace{-0.4cm} 3.9 $\pm$ 1.4 & 0.81 $\pm$ 0.12 & --- $\pm$ --- & \cite {2017AJ....154..205M} \\
\hspace{-0.9cm} OGLE-2013-BLG-1721 & \hspace{-0.4cm}$ 245.0 ^{+ 501.0 } _{- 146.0 }$ & \hspace{-0.4cm} 6.3 $\pm$ 1.35 & 0.42 $\pm$ 0.09 & --- $\pm$ --- & \cite {2017AJ....154..205M} \\
\hspace{-0.9cm} OGLE-2014-BLG-0124 & \hspace{-0.4cm}$ 165.0 ^{+ 145.0 } _{- 91.0 }$ & \hspace{-0.4cm} 4.1 $\pm$ 0.59 & 0.84 $\pm$ 0.26 & --- $\pm$ --- & \cite {2015ApJ...799..237U} \\
\hspace{-0.9cm} OGLE-2014-BLG-0676 & \hspace{-0.4cm}$ 1010.0 ^{+ 1230.0 } _{- 640.0 }$ & \hspace{-0.4cm} 2.22 $\pm$ 0.895 & 1.38 $\pm$ 0.43 & 1.81 $\pm$ 0.44 & \cite {2017MNRAS.466.2710R} \\
\hspace{-0.9cm} OGLE-2014-BLG-1760 & \hspace{-0.4cm}$ 63.0 ^{+ 103.0 } _{- 31.0 }$ & \hspace{-0.4cm} 6.86 $\pm$ 1.11 & 0.29 $\pm$ 0.05 & --- $\pm$ --- & \cite {2016AJ....152..140B} \\
\hspace{-0.9cm} OGLE-2015-BLG-0051 & \hspace{-0.4cm}$ 286.0 ^{+ 759.0 } _{- 166.0 }$ & \hspace{-0.4cm} 8.2 $\pm$ 0.9 & 0.093 $\pm$ 0.008 & --- $\pm$ --- & \cite {2016AJ....152...95H} \\
\hspace{-0.9cm} OGLE-2015-BLG-0954 & \hspace{-0.4cm}$ 1096.0 ^{+ 646.0 } _{- 545.0 }$ & \hspace{-0.4cm} 0.6 $\pm$ 0.3 & 1.89 $\pm$ 0.1701 & --- $\pm$ --- & \cite {2016JKAS...49...73S} \\
\hspace{-0.9cm} OGLE-2015-BLG-0966 & \hspace{-0.4cm}$ 21.12 ^{+ 3.5 } _{- 3.37 }$ & \hspace{-0.4cm} 2.5 $\pm$ 0.0 & 0.76 $\pm$ 0.07 & --- $\pm$ --- & \cite {2016ApJ...819...93S} \\
\hspace{-0.9cm} OGLE-2016-BLG-0613 & \hspace{-0.4cm}$ 1850.0 ^{+ 2220.0 } _{- 1090.0 }$ & \hspace{-0.4cm} 3.41 $\pm$ 1.38 & 1.2 $\pm$ 0.24 & --- $\pm$ --- & \cite {2017AJ....154..133H} \\
\hspace{-0.9cm} OGLE-2016-BLG-1190 & \hspace{-0.4cm}$ 4249.0 ^{+ 561.0 } _{- 478.0 }$ & \hspace{-0.4cm} 6.77 $\pm$ 0.085 & 0.49 $\pm$ 0.04 & --- $\pm$ --- & \cite {2018AJ....155...40R} \\
\hspace{-0.9cm} OGLE-2016-BLG-1195 & \hspace{-0.4cm}$ 1.37 ^{+ 0.66 } _{- 0.48 }$ & \hspace{-0.4cm} 3.91 $\pm$ 0.44 & 0.286 $\pm$ 0.0455 & 0.4 $\pm$ 0.1 & \cite {2017ApJ...840L...3S} \\
\hspace{-0.9cm} OGLE-2017-BLG-0173 & \hspace{-0.4cm}$ 3.3 ^{+ 4.94 } _{- 1.84 }$ & \hspace{-0.4cm} 4.705 $\pm$ 1.6155 & 0.54 $\pm$ 0.03 & --- $\pm$ --- & \cite {2018AJ....155...20H} \\
\hspace{-0.9cm} TCP J05074264+2447555 & \hspace{-0.4cm}$ 7.27 ^{+ 2.94 } _{- 2.35 }$ & \hspace{-0.4cm} 0.38 $\pm$ 0.0 & 1.45 $\pm$ 0.25 & --- $\pm$ --- & \cite {2018MNRAS.476.2962N} \\
\hspace{-0.9cm} UKIRT-2017-BLG-001 & \hspace{-0.4cm}$ 340.0 ^{+ 570.0 } _{- 209.0 }$ & \hspace{-0.4cm} 6.3 $\pm$ 1.85 & 0.91 $\pm$ 0.475 & --- $\pm$ --- & \cite {2018arXiv180206795S} \\
\hspace{-0.9cm} MOA-2010-BLG-117 & \hspace{-0.4cm}$ 168.6 ^{+ 76.4 } _{- 52.7 }$ & \hspace{-0.4cm} 3.5 $\pm$ 0.4 & 0.777 $\pm$ 0.095 & --- $\pm$ --- & \cite {2018AJ....155..141B} \\
\hspace{-0.9cm} OGLE-2017-BLG-1522 & \hspace{-0.4cm}$ 187.0 ^{+ 429.0 } _{- 104.0 }$ & \hspace{-0.4cm} 7.49 $\pm$ 0.895 & 0.065 $\pm$ 0.009 & --- $\pm$ --- & \cite {2018arXiv180305095J} \\
\hline  
\end{tabular}
\end{table}

\end{document}